\documentstyle[twoside,axodraw,psfig]{article}

\catcode`\@=11
\long\def\@makefntext#1{
\protect\noindent \hbox to 3.2pt {\hskip-.9pt  
$^{{\eightrm\@thefnmark}}$\hfil}#1\hfill}		

\def\@makefnmark{\hbox to 0pt{$^{\@thefnmark}$\hss}}	
	
\def\ps@myheadings{\let\@mkboth\@gobbletwo
\def\@oddhead{\hbox{}
\rightmark\hfil\eightrm\thepage}   
\def\@oddfoot{}\def\@evenhead{\eightrm\thepage\hfil
\leftmark\hbox{}}\def\@evenfoot{}
\def\sectionmark##1{}\def\subsectionmark##1{}}

\oddsidemargin=\evensidemargin
\newcounter{sectionc}\newcounter{subsectionc}\newcounter{subsubsectionc}
\renewcommand{\section}[1] {\vspace{12pt}\addtocounter{sectionc}{1} 
\setcounter{subsectionc}{0}\setcounter{subsubsectionc}{0}\noindent 
	{\tenbf\thesectionc. #1}\par\vspace{5pt}}
\renewcommand{\subsection}[1] {\vspace{12pt}\addtocounter{subsectionc}{1} 
	\setcounter{subsubsectionc}{0}\noindent 
	{\bf\thesectionc.\thesubsectionc. {\kern1pt \bfit #1}}\par\vspace{5pt}}
\renewcommand{\subsubsection}[1] {\vspace{12pt}\addtocounter{subsubsectionc}{1}
	\noindent{\tenrm\thesectionc.\thesubsectionc.\thesubsubsectionc.
	{\kern1pt \tenit #1}}\par\vspace{5pt}}
\newcommand{\nonumsection}[1] {\vspace{12pt}\noindent{\tenbf #1}
	\par\vspace{5pt}}

\newcounter{appendixc}
\newcounter{subappendixc}[appendixc]
\newcounter{subsubappendixc}[subappendixc]
\renewcommand{\thesubappendixc}{\Alph{appendixc}.\arabic{subappendixc}}
\renewcommand{\thesubsubappendixc}
	{\Alph{appendixc}.\arabic{subappendixc}.\arabic{subsubappendixc}}

\renewcommand{\appendix}[1] {\vspace{12pt}
        \refstepcounter{appendixc}
        \setcounter{figure}{0}
        \setcounter{table}{0}
        \setcounter{lemma}{0}
        \setcounter{theorem}{0}
        \setcounter{corollary}{0}
        \setcounter{definition}{0}
        \setcounter{equation}{0}
        \renewcommand{\thefigure}{\Alph{appendixc}.\arabic{figure}}
        \renewcommand{\thetable}{\Alph{appendixc}.\arabic{table}}
        \renewcommand{\theappendixc}{\Alph{appendixc}}
        \renewcommand{\thelemma}{\Alph{appendixc}.\arabic{lemma}}
        \renewcommand{\thetheorem}{\Alph{appendixc}.\arabic{theorem}}
        \renewcommand{\thedefinition}{\Alph{appendixc}.\arabic{definition}}
        \renewcommand{\thecorollary}{\Alph{appendixc}.\arabic{corollary}}
        \renewcommand{\theequation}{\Alph{appendixc}.\arabic{equation}}
        \noindent{\tenbf Appendix \theappendixc #1}\par\vspace{5pt}}
\newcommand{\subappendix}[1] {\vspace{12pt}
        \refstepcounter{subappendixc}
        \noindent{\bf Appendix \thesubappendixc. {\kern1pt \bfit #1}}
	\par\vspace{5pt}}
\newcommand{\subsubappendix}[1] {\vspace{12pt}
        \refstepcounter{subsubappendixc}
        \noindent{\rm Appendix \thesubsubappendixc. {\kern1pt \tenit #1}}
	\par\vspace{5pt}}

\newcommand{\textlineskip}{\baselineskip=13pt}
\newcommand{\smalllineskip}{\baselineskip=10pt}

\def\eightcirc{
\begin{picture}(0,0)
\put(4.4,1.8){\circle{6.5}}
\end{picture}}
\def\eightcopyright{\eightcirc\kern2.7pt\hbox{\eightrm c}} 

\newcommand{\copyrightheading}[1]
	{\vspace*{-2.5cm}\smalllineskip{\flushleft
	{\footnotesize International Journal of Modern Physics A #1}\\
	{\footnotesize $\eightcopyright$\, World Scientific Publishing
	 Company}\\
	 }}

\newcommand{\publisher}[2]{{\begin{center}\footnotesize\smalllineskip 
	Received #1\\
	Revised #2
	\end{center}
	}}

\def\abstracts#1#2#3{{
	\centering{\begin{minipage}{4.5in}\footnotesize\baselineskip=10pt
	\parindent=0pt #1\par 
	\parindent=15pt #2\par
	\parindent=15pt #3
	\end{minipage}}\par}} 

\newcommand{\bibit}{\nineit}

\renewenvironment{thebibliography}[1]
	{\frenchspacing
	 \ninerm\baselineskip=11pt
	 \begin{list}{\arabic{enumi}.}
	{\usecounter{enumi}\setlength{\parsep}{0pt}
	 \setlength{\leftmargin 12.7pt}{\rightmargin 0pt} 
	 \setlength{\itemsep}{0pt} \settowidth
	{\labelwidth}{#1.}\sloppy}}{\end{list}}

\newcounter{itemlistc}
\newcounter{romanlistc}
\newcounter{alphlistc}
\newcounter{arabiclistc}

\newcommand{\fcaption}[1]{
        \refstepcounter{figure}
        \setbox\@tempboxa = \hbox{\footnotesize Fig.~\thefigure. #1}
        \ifdim \wd\@tempboxa > 5in
           {\begin{center}
        \parbox{5in}{\footnotesize\smalllineskip Fig.~\thefigure. #1}
            \end{center}}
        \else
             {\begin{center}
             {\footnotesize Fig.~\thefigure. #1}
              \end{center}}
        \fi}

\newcommand{\tcaption}[1]{
        \refstepcounter{table}
        \setbox\@tempboxa = \hbox{\footnotesize Table~\thetable. #1}
        \ifdim \wd\@tempboxa > 5in
           {\begin{center}
        \parbox{5in}{\footnotesize\smalllineskip Table~\thetable. #1}
            \end{center}}
        \else
             {\begin{center}
             {\footnotesize Table~\thetable. #1}
              \end{center}}
        \fi}

\def\@citex[#1]#2{\if@filesw\immediate\write\@auxout
	{\string\citation{#2}}\fi
\def\@citea{}\@cite{\@for\@citeb:=#2\do
	{\@citea\def\@citea{,}\@ifundefined
	{b@\@citeb}{{\bf ?}\@warning
	{Citation `\@citeb' on page \thepage \space undefined}}
	{\csname b@\@citeb\endcsname}}}{#1}}

\newif\if@cghi
\def\cite{\@cghitrue\@ifnextchar [{\@tempswatrue
	\@citex}{\@tempswafalse\@citex[]}}
\def\citelow{\@cghifalse\@ifnextchar [{\@tempswatrue
	\@citex}{\@tempswafalse\@citex[]}}
\def\@cite#1#2{{$\null^{#1}$\if@tempswa\typeout
	{IJCGA warning: optional citation argument 
	ignored: `#2'} \fi}}

\def\pmb#1{\setbox0=\hbox{#1}
	\kern-.025em\copy0\kern-\wd0
	\kern.05em\copy0\kern-\wd0
	\kern-.025em\raise.0433em\box0}

\def\fnt#1#2{\footnotetext{\kern-.3em
	{$^{\mbox{\scriptsize #1}}$}{#2}}}

\def\@makefnmark{\hbox to 0pt{$^{\@thefnmark}$\hss}}	
	
\def\ps@myheadings{%
    \let\@oddfoot\@empty\let\@evenfoot\@empty
    \def\@evenhead{\slshape\leftmark\hfil}
    \def\@oddhead{\hfil{\slshape\rightmark}}
    \let\@mkboth\@gobbletwo
    \let\sectionmark\@gobble
    \let\subsectionmark\@gobble
    }
\font\tenrm=cmr10
\font\tenit=cmti10 
\font\tenbf=cmbx10
\font\bfit=cmbxti10 at 10pt
\font\ninerm=cmr9
\font\nineit=cmti9

\font\eightrm=cmr8

\def\qed{\hbox{${\vcenter{\vbox{			
   \hrule height 0.4pt\hbox{\vrule width 0.4pt height 6pt
   \kern5pt\vrule width 0.4pt}\hrule height 0.4pt}}}$}}

\begin{document}
\setlength{\textheight}{7.7truein}  

\thispagestyle{empty}

\markboth{\protect{\footnotesize\it Photon Propagation in a Magnetized Medium}}
{\protect{\footnotesize\it Photon Propagation in a Magnetized Medium}}

\normalsize\textlineskip

\setcounter{page}{1}

\copyrightheading{}		

\vspace*{0.88truein}

\centerline{\bf PHOTON PROPAGATION IN A MAGNETIZED MEDIUM}
\vspace*{0.37truein}
\centerline{\footnotesize SUSHAN KONAR$^{*}$}
\baselineskip=12pt
\centerline{\footnotesize\it Inter-University Centre for Astrophysics \& Astronomy}
\baselineskip=10pt
\centerline{\footnotesize\it Pune, Maharashtra 411007, India}
\baselineskip=10pt
\centerline{\footnotesize\it $^{*}$E-mail: sushan@iucaa.ernet.in}
\vspace*{0.225truein}
\publisher{(received date)}{(revised date)}

\vspace*{0.21truein}
\abstracts{Using the real time formalism of the finite temperature field theory we calculate the 1-loop 
polarization tensor in the presence of a background magnetic field in a medium. The expression is obtained
to linear order in the background field strength. We discuss the Faraday rotation as well as the photon 
absorption probabilities in this context.} {}{}


\vspace*{1pt}\textlineskip	
\section{Introduction}	        
\vspace*{-0.5pt}
\noindent
The propagation of electro-magnetic waves, in a magnetised plasma, is of interest in systems ranging from 
laboratory plasma to astrophysical objects~\cite{1}. Yet, the expression for Faraday Rotation, for example, 
is derived assuming the medium to consist of non-relativistic and non-degenerate particles. Since, such
assumptions may not be valid in every context we re-investigate this problem in a general framework. 

Since, almost all the physical systems have magnetic fields smaller than the QED limit ($eB < m_e^2$) a 
weak-field treatment is justified. Moreover, in compact astrophysical objects (white dwarfs/neutron stars) 
the Landau level spacings are negligible compared to the electron Fermi energy~\cite{2}. Hence, we can also 
assume that the field does not introduce any spatial anisotropy in the collective plasma behaviour. 
 
Therefore, we calculate the polarization tensor ($\Pi_{\mu \nu}$), at the 1-loop level, in the weak-field limit 
retaining terms up-to ${\cal O(B)}$. As expected, we recover Faraday rotation from the dispersive part of the 
polarization tensor and the absorptive part provides the damping/instability of the photons propagating in a 
plasma. These calculations have already been reported in detail in two recent articles by us (1999, 2001)~\cite{3,4}. 

\section{The Formalism}
\noindent The presence of an external field or a medium introduces quantum corrections to the Lagrangian of an 
electro-magnetic field. In the momentum space the quadratic part of the Lagrangian, inclusive of such corrections, 
is given by,
\begin{equation}
{\cal L} = \frac{1}{2} [-k^2 (g_{\mu \nu} - \frac{k_\mu k_\nu}{k^2}) + \Pi_{\mu \nu}(k)] A^\mu(k) A^\nu(k) \,,
\end{equation}
where $\Pi_{\mu \nu}(k)$ is the polarization tensor. Assuming the direction of photon propagation to be along 
the direction of the magnetic field, the dispersion relation for the two transverse components of the photon field 
$A^{\mu}$, is:
\begin{equation}
k^2 = \omega_0^2 \pm \left( a_{\rm disp} + a_{\rm abs} \right) \,,
\label{disp1}
\end{equation}
where $a_{\rm disp}$ and $a_{\rm abs}$ are the dispersive and the absorptive parts of $\Pi_{\mu \nu}$.
Here we assume the Lorenz gauge condition ($\partial_\mu A^{\mu} = 0$) and $\omega_0$ is the plasma frequency.
For a plane polarized electro-magnetic wave propagating with a frequency $\omega$ the rate of rotation of the 
polarization angle, per unit length $l$, (i.e, the Faraday Rotation) is then given by:
\begin{equation}
\frac{d\Phi}{dl} = \frac{a_{\rm disp}}{\sqrt{\omega^2 - \omega_0^2}} \,.
\end{equation}

\section{1-Loop Vacuum Polarization}

\begin{figure}
\begin{center}
\begin{picture}(150,50)(0,-25)
\Photon(0,0)(40,0){2}{4}
\Text(20,5)[b]{$k\rightarrow$}
\Photon(110,0)(150,0){2}{4}
\Text(130,5)[b]{$k\rightarrow$}
\Text(75,30)[b]{$p+k\equiv p'$}
\Text(75,-30)[t]{$p$}
\SetWidth{1.2}
\Oval(75,0)(25,35)(0)
\ArrowLine(74,25)(76,25)
\ArrowLine(76,-25)(74,-25)
\end{picture}
\end{center}
\caption[]{One-loop diagram for the vacuum polarization.}\label{f:1loop}
\end{figure}
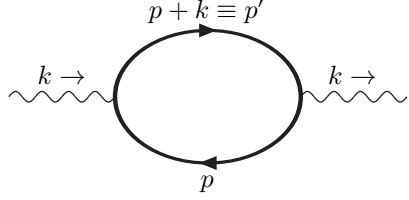
\noindent At the 1-loop level, the polarization tensor arises from the diagram in fig.~\ref{f:1loop},
with the dominant contribution coming from the electron line in the loop. To evaluate this diagram we use the 
electron propagator within a thermal medium in presence of a background electro-magnetic field. Since, we
specialize to the case of a purely magnetic field, the field can be taken to be in the $z$-direction 
without any loss of generality. The magnitude of this field is denoted by $\cal B$. In presence of such a
background field, the vacuum electron propagator is (following Schwinger~\cite{5}):
\begin{equation}
i S_B^V(p) = \int_0^\infty ds\; e^{\Phi(p,s)} C(p,s) \,,
\label{SV2}
\end{equation}
using the shorthands,
\begin{eqnarray}
\Phi(p,s) &\equiv& is \left( p_\parallel^2 - {\tan (e{\cal B}s) \over e{\cal B}s}
\, p_\perp^2 - m^2 \right) - \epsilon |s| \,,
\label{Phi} \\
C(p,s) &\equiv& \Big[ ( 1 + i\sigma_z \tan  e{\cal B}s ) (\rlap/p_\parallel + m )
- (\sec^2 e{\cal B}s) \rlap/ p_\perp \Big] \,.
\label{C}
\end{eqnarray}
where, $\sigma_z = i\gamma_1 \gamma_2$. To write $\Phi(p,s)$ and $C(p,s)$ we have used the following decomposition 
of the metric tensor : $g_{\mu \nu} = g^{\parallel}_{\mu \nu} - g^{\perp}_{\mu \nu}$, where, 
$g^{\parallel}_{\mu \nu}$~= diag(1,0,0,-1) and $g^{\perp}_{\mu \nu}$~= diag(0,1,1,0). In the presence of a background 
medium, the above propagator is modified to~\cite{6}:
\begin{equation}
iS(p) = iS_{\rm B}^{\rm V}(p) + S^{\eta}_{\rm B}(p) \,,
\label{fullprop}
\end{equation}
where $S^{\eta}_{\rm B}(p) =  - \eta_F(p) \left[ iS_{\rm B}^{\rm V}(p) - i\overline S_{\rm B}^{\rm V}(p) \right]$ and 
$\overline S_{\rm B}^{\rm V}(p) \equiv \gamma_0 S^{{\rm V} \dagger}_{\rm B}(p) \gamma_0$ for a fermion propagator. 
And $\eta_F(p)$ contains the distribution function for the fermions and the anti-fermions, given by:
\begin{equation}
\eta_F(p) = \Theta(p\cdot u) f_F(p,\mu,\beta) + \Theta(-p\cdot u) f_F(-p,-\mu,\beta) \,,
\label{eta}
\end{equation}
where $u$ and $\beta$ are the 4-velocity and the effective temperature of the background thermal medium. Here, 
$f_F(p,\mu,\beta) = (e^{\beta(p\cdot u - \mu)} + 1)^{-1}$ is the Fermi-Dirac distribution function and $\Theta$ is 
the step function. Rewriting eq.(\ref{fullprop}) in the following form:
\begin{equation}
iS(p) = \frac{i}{2} \left[ S_B^V(p) + \overline S_B^V(p) \right] + i (1/2 - \eta_F(p)) \left[ S_B^V(p) - \overline S_B^V(p) \right] \,
\label{S_reim}
\end{equation}
we recognise:
\begin{equation}
S_{\rm re} =  \frac{1}{2} \left[ S_B^V(p) + \overline S_B^V(p) \right] \; , \;\; 
S_{\rm im} = (1/2 - \eta_F(p)) \left[ S_B^V(p) - \overline S_B^V(p) \right] \,.
\end{equation}
The subscripts {\em re} and {\em im} refer to the real and imaginary parts of the propagator. Now, the amplitude of the 
1-loop diagram of fig.~\ref{f:1loop} can be written as:
\begin{eqnarray}
i \Pi_{\mu\nu}(k) = - \int \frac{d^4p}{(2\pi)^4} (ie)^2 \; \mbox{tr}\,
\left[\gamma_\mu \, iS(p) \gamma_\nu \, iS(p')\right] \,,
\end{eqnarray}
where, for the sake of notational simplicity, we have used $p' = p+k$. Then the dispersive part of the polarisation 
tensor is given by:
\begin{equation}
\Pi^{\rm D}_{\mu\nu}(k) = -ie^2 \int \frac{d^4p}{(2\pi)^4} \; \mbox{tr}\,
\left[\gamma_\mu \, S^{\eta}_{\rm B}(p) \gamma_\nu \, iS^{\rm V}_{\rm B}(p^{\prime}) + 
\gamma_\mu \, iS^{\rm V}_{\rm B}(p) \gamma_\nu \, iS^{\eta}_{\rm B}(p^{\prime}) \right] \,,       
\end{equation}
and the (11)-component of the absorptive part, by:
\begin{equation}
\Pi_{\mu\nu}^{A}(k) = -ie^2 \int \frac{d^4p}{(2\pi)^4} \; \mbox{tr}\,
\left[\gamma_\mu \, iS_{\rm im}(p) \gamma_\nu \, iS_{\rm im}(p^{\prime}) \right]\,.        
\end{equation}
Now, for the terms odd in powers of ${\cal B}$, the explicitly gauge invariant expressions are (see~\cite{3,4} for details):
\begin{eqnarray}
\Pi^{\rm D}_{\mu\nu}(k) 
&=& 4ie^2 \varepsilon_{\mu \nu \alpha_{\parallel} \beta} k^\beta 
    \int \frac{d^4p}{(2\pi)^4} \; \eta_-(p)
    \int_{-\infty}^\infty ds e^{\Phi(p,s)} \int_0^\infty ds' e^{\Phi(p',s')} \nonumber \\
&\times&  \left[ p^{\widetilde{\alpha_\parallel}} \tan e{\cal B}s + p^{\prime \widetilde{\alpha_\parallel}} \tan e{\cal B}s' 
    - \frac{\tan e{\cal B}s \tan e{\cal B}s'}{\tan e{\cal B}(s+s')} (p + p')^{\widetilde{\alpha_\parallel}} \right] \,,       
\end{eqnarray}
and,
\begin{eqnarray}
\Pi^{\rm A}_{\mu\nu}(k) 
&=& - \, ie^2 \varepsilon_{\mu\nu\alpha_\parallel\beta} k^\beta \, \int \frac{d^4p}{(2\pi)^4} X(\beta, k, p) 
      \int_{-\infty}^\infty ds \; e^{\Phi(p,s)} \int_{-\infty}^\infty ds' \; e^{\Phi(p',s')} \nonumber\\* 
&\times& \Bigg[ p^{\widetilde\alpha_\parallel} \tan e{\cal B}s + p'^{\widetilde\alpha_\parallel} \tan e{\cal B}s' 
       - {\tan e{\cal B}s \; \tan e{\cal B}s' \over \tan e{\cal B}(s+s')} \; (p+p')^{\widetilde\alpha_\parallel} \Bigg] \,. 
\end{eqnarray}
In writing the above expressions we have used the notation of 
$p^{\widetilde{\alpha_\parallel}}$, for example. This signifies a component of $p$ which can take only the "parallel" 
indices, i.e., 0 or 3 and is moreover different from the index $\alpha$ appearing elsewhere in the expression. We have 
also used the shorthands, $\eta_-(p) = \eta_F(p) - \eta_F(-p)$ and 
$X(\beta, k, p) = (1 - 2 \eta_F(p)) \, (1 - 2 \eta_F(p^{\prime}))$.

\section{Results}

\noindent Our results are obtained in the rest frame of the background medium. We also take the long wavelength limit,
i.e, $K \ll \omega$, where ($k_0 = \omega$). Finally, we assume the magnetic field to be small such that we can retain 
terms only linear in ${\cal B}$. In this limit, the dispersive part of the polarization tensor is given by:
\begin{eqnarray}
\Pi^{\rm D}_{\mu\nu}(k) 
&=& 8ie^2 \varepsilon_{\mu \nu \alpha_{\parallel} \beta} {\cal B} \omega 
    \int \frac{d^4p}{(2\pi)^4} \; \eta_-(p) p_0
    \int_{-\infty}^\infty ds e^{is(p^2 - m^2) - \epsilon|s|} \nonumber \\
&\times& \int_0^\infty ds' e^{is(p'^2 - m'^2) - \epsilon|s'|} \left\{s+s' - \frac{ss'}{s+s'}\right\},
\end{eqnarray}
where we have made a further assumption that $\omega \ll m_e$ (see~\cite{3} for details). Surprisingly, in the above 
mentioned limit, the absorptive part of the polarization tensor has two terms with different signs which opens up the 
possibility that for a given magnetic field, depending on the chemical potential and the external photon momentum 
$\Pi^{\rm A}_{\mu \nu}$ can be either positive or negative giving rise to damping or instability of the propagating 
photon (see~\cite{4} for details). Also, the absorption of the photons happen between two limiting values of $p$ given by,
\begin{eqnarray}
P_{\rm min} =  - \frac{K}{2} + \frac{k_0}{2} \left(1- \frac{4 m^2}{k^2}\right)^{1/2}, \; \;
P_{\rm max} =  \frac{K}{2} + \frac{k_0}{2} \left(1- \frac{4 m^2}{k^2}\right)^{1/2}.
\end{eqnarray}
Since, $P$ is real, the condition $k^2 \ge 4 m^2$ must be satisfied. This is an important kinematic constraint which ensures
the conservation of energy-momentum in the weak-field limit.

\nonumsection{Acknowledgments}
\noindent
Participation to this conference was made possible through a travel grant (No. TG/390/01-HRD) 
from the Council of Scientific and Industrial Research of India.

\nonumsection{References}

\end{document}